# Performance of Social Network Sensors During Hurricane Sandy

Yury Kryvasheyeu[1], Haohui Chen[1], Esteban Moro[2,3], Pascal Van Hentenryck[1,4], and Manuel Cebrian[1,*]

[1]*National Information and Communications Technology Australia, University of Melbourne, Victoria 3010, Australia*
[2]*Department of Mathematics & GISC, Universidad Carlos III de Madrid, Madrid, Leganés 28911, Spain*
[3]*Instituto de Ingeniería del Conocimiento, Universidad Autónoma de Madrid, Madrid 28049, Spain*
[4]*Research School of Computer Science, Australian National University, ACT 0200, Australia*

*\* Corresponding Author: MC, manuel.cebrian@nicta.com.au*

**Information flow during catastrophic events is a critical aspect of disaster management. Modern communication platforms, in particular online social networks, provide an opportunity to study such flow and derive early-warning sensors, thus improving emergency preparedness and response. Performance of the social networks sensor method, based on topological and behavioural properties derived from the "friendship paradox", is studied here for over 50 million Twitter messages posted before, during, and after Hurricane Sandy. We find that differences in users' network centrality effectively translate into moderate awareness advantage (up to 26 hours); and that geo-location of users within or outside of the hurricane-affected area plays a significant role in determining the scale of such an advantage. Emotional response appears to be universal regardless of the position in the network topology, and displays characteristic, easily detectable patterns, opening a possibility to implement a simple "sentiment sensing" technique that can detect and locate disasters.**

# 1   Introduction

Natural, man-made and technological disasters present a constant threat to society [1]. An increasing frequency, intensity and impact of such events are often attributed to the effects of climate change [2-4]. Consensus is growing that the likelihood and potential damage of natural disasters in the future will rise, and there is a need to adequately prepare for their consequences [5-8]. An integral part of such preparation efforts is an understanding of the information flow during disasters in order to derive early-warning sensors, track public awareness, gather emergency and relief information and predict human behaviour, such as escape panic [9-11]. Conveniently, online social media, like Twitter and Facebook, have matured into prominent communication platforms and provide an unprecedented opportunity to record and analyse vast amounts of information [12]. The potential of these networks is already leveraged during natural disasters [13-17], with applications in situation awareness [18,19], event detection [20,21], search/locating persons [22,23] or helping through crowdsourcing initiatives [24,25].

One particular network phenomenon, the "friendship paradox", may increase efficiency of monitoring disaster related information. The paradox was studied in a seminal work by Feld [26] and known colloquially as "your friends have more friends than you do". Feld showed that a node in a social network on average has lower number of links than the average of links its friends have. This occurs because well-connected nodes are included multiple times in a set of "friends-of-friends", therefore boosting corresponding average. The paradox and its strong form (formulated for median rather than mean averaging) were observed experimentally in the context of online social networks [27,28] and networks of co-authorships and citations [29]. Although the original focus of Feld's study was on psychological implications of the paradox (e.g. potential perception of inadequate social inclusion), his finding inspired a simple technique of forming a sample group with network centrality above what random sampling allows. This could be achieved without global knowledge of network topology by using friends of randomly selected people instead of themselves. Because centrality often appears in correlation with other attributes – like activity, popularity, health or income [27,29] – friends are subject to an earlier exposure to a contagion [30,31] or information propagating through the network [32].

In a disaster, the ability to implement efficient and early detection of emergency information is extremely valuable and a sensor method technique based on the "friendship paradox" is attractive for that purpose. Existing experimental validations of the sensor method [30,32] confirm its applicability for endogenous processes, when the spread of contagion or exchange of information occurs only between the nodes of a network. Contrary to such a spread of infection or social network memes, information about disasters is carried simultaneously by many other external channels. Considering this interplay of exogenous and endogenous propagation modes, and factoring in the speed, scale and strong geographical nature of phenomena such as hurricanes, it is not immediately obvious whether the sensor method would perform reliably in a disaster. To address this central question, we study performance of the sensor method during Hurricane Sandy to establish if there is an early awareness advantage, what is its magnitude, and what is the effect of geographical location of users. Finally, while there is evidence that centrality correlates with measurable attributes [27-29], it is unclear if there is an underlying correlation with personality or behaviour traits. To study this, we employ sentiment analysis to explore differences between random control groups and corresponding sensor groups in terms of the timeliness and magnitude of their emotional response.

## 2 Data and Methods

*2.1 Context of the research: Hurricane Sandy and its digital traces on Twitter*

The disaster event at the centre of our case study is Hurricane Sandy, the largest hurricane of the 2012 season and one of the costliest disasters in the history of the United States. Sandy was a late season hurricane that formed on October 22 2012 at 12:00 UTC about 500 km south-southwest of Kingston, Jamaica. It made its first landfall in Jamaica at 19:00 UTC on October 24 as a Category 1 hurricane, then as a Category 3 hurricane in Cuba at 05:30 UTC on October 25, subsequently weakening down to Category 1 as it moved through the Bahamas. It continued to grow in size and, while moving northeast along the United States coast, re-intensified to the maximum wind speeds of 85 knots at 12:00 UTC on October 29, about 350 km southeast of Atlantic City. In the next day the hurricane weakened into a post-tropical storm and made its landfall at 23:30 UTC on October 29 near Brigantine in New Jersey. At the time of landfall the wind reached 70 knots and the storm surge was as high as 3.85 meters, with prevalent levels between 0.8 and 2.8 meters along the coast of New Jersey and New York. The storm surge was responsible for most of the damage to houses, totalling up to 650,000 destroyed or damaged buildings. Over 8.5 million people were affected by power losses that lasted for weeks in some areas. According to the National Hurricane Center report [33], Sandy caused 147 direct casualties along its path and brought damage in excess of $50 billion for the United States.

## 2.2 Raw datasets

Hurricane Sandy attracted extensive media coverage, both in traditional broadcasting media and online. In our work we look at digital traces of the hurricane on Twitter. Raw data collected for analysis is comprised of two principal sets of Twitter messages. The first one consists of messages with the hashtag "#sandy" posted between October 15 and November 12, in the period that precedes the formation of the hurricane and extends beyond its landfall in the continental United States. The data includes the text of messages and a range of additional information, such as message identifiers, user identifiers, followees counts, re-tweet statuses, self-reported or automatically detected locations, timestamps, and sentiment scores. The second dataset has similar structure and is collected within the same timeframe; however, instead of a hashtag, it includes all messages that contain one or more instances of specific keywords, deemed to be relevant to the event and its consequences ("sandy", "hurricane", "storm", "superstorm", "flooding", "blackout",

"gas", "power", "weather", "climate", etc.). The full list of keywords used to build up the dataset is provided in the Supplementary Information, Table S6. In total there are 52.55 million messages from 13.75 million unique users available for analysis.

## 2.3 Location data and geocoding

Raw data is filtered to include only those messages that contain location information. Since only a minor fraction of the messages (about 1.2% for the hashtag dataset and 1.5% for the keywords dataset) are geo-tagged by Twitter, we attempt to extract additional information from incomplete self-reported data in user profiles. Such data includes self-identification of a country, state, province, city or town, or any arbitrary combination of those items. We analyse only profile data, instead of searching for location-specific text within messages, to avoid the ambiguity of dealing with context of such in-text location mentions (hypothetical, past or future travels, messages about other people, abstract mentions of various places, etc.). After crosschecking partial location information against coordinates of all major administrative regions and cities worldwide, 46% of messages (or 43% of users) were encoded with location data. Precision of geocoding varies between the exact latitude and longitude of a user, as recorded by Twitter, and the coordinates of the centre of an administrative unit that returned a match to a self-reported location. The rate of location detection compares favourably with other studies, e.g. 6.6% detection rate in Mislove et al. [34], where Google Maps API interpretation of self–reported location strings was used to obtain coordinates.

We further filter for users from the United States and Canada, to reduce variations in time zones and languages of tweets. Hurricane path and extent data, shown in Figure 1, are utilised to distinguish users based on whether they were affected by Sandy directly. After filtering for geocoded messages the hashtag dataset includes 3.65 million messages from 1.24 million users (25.6% of them directly affected by the hurricane) and the keyword dataset includes 24.15 million messages from 5.98 million users (14.1% in the affected region).

## 2.4 Relevance filtering

The last filter that we impose on the raw data is content analysis to insure a message is relevant to Hurricane Sandy. Our study of the awareness advantage relies on the time of the first hurricane-related tweet for each user, which must be determined as correctly as possible. Potential issues may arise equally from data incompleteness or excess. Incomplete data is a problem for the hashtag-based dataset, because hashtags are not used systematically in every message and some (or even all) relevant messages may be overlooked. In our case, the hashtag dataset includes 3.65 million messages, but the same users within the same period of time are represented by 11.07 million messages in the extended dataset. Although some additional messages are not related to the hurricane, many are, and must be included in the analysis. To avoid "false positives" (messages with no relevance to Hurricane Sandy) we implement simple filtering described below.

The evolution of Hurricane Sandy provides a convenient frame of reference in order to check the relevance. Since the hurricane was first classified as such and officially assigned its name on October 22, any keyword with a significant level of activity before that date should be filtered out to avoid inclusion of unrelated information. Figures S1 and S2 of the Supplementary Information summarize the histograms of messages matching specific keywords and demonstrate that the majority of them suffer from irrelevance noise. Regrettably, certain keywords of interest ("storm", "power" and "gas") are contaminated by irrelevant messages simply due to their general nature and/or multiple meanings: for instance "storm" is mentioned in messages about small scale local weather events and "power" is used not just in the context of post-hurricane power outages, but also in context of politics against a backdrop of the presidential election campaign. To eliminate noise from the datasets, only messages with a word "sandy", either in a hashtag or keyword form, were included in the analysis. The effect of filtering is demonstrated in Figure 2, which compares histograms for messages without filtering, moderate filtering (words include "sandy", "storm", "hurricane", "huracán", "superstorm" and "frankenstorm"), and strict filtering ("sandy"). The results indicate that only the strict filtering succeeds in suppressing noise messages prior to the formation of the hurricane. Relevance filtering brings the total volume of data down to 4.51 million messages from 1.39 million unique users.

## 3 Results

### 3.1 Lead-time in awareness

Arguably the most important aspect of the information flow during a disaster is awareness. Given the limitations of our dataset and the lack of retrospective studies into the link between disaster awareness and patterns of online activity, we assume that the time a person becomes aware about the hurricane and tweets about it are close to each other. To evaluate performance of the sensor method, we focus on the entry time $t$, defined as the time a user first appears in our dataset by posting a message relevant to Hurricane Sandy. Following the conventional terminology, we call an original random sample a control group, and a group formed from their friends a sensor group. Let us define the lead-time as the difference between the average entry times of the sensor group and its corresponding control group: $\Delta t = \langle t_S \rangle - \langle t_C \rangle$, with negative lead-times indicating awareness advantage. We estimate lead-times of sensor over control groups across the range of sample sizes from 500 to 100,000 users. Control groups are formed by random selection from the pool of users with known geographical location. Sensor groups are formed from the friends (followees) of users in their corresponding control groups; the two groups are of the same size and without user duplication. In all the analysis reported below and in the horizontal axes of figures, times are given as offsets in hours with respect to the 00:00 UTC on October 30, which is approximately the time of Sandy's landfall on the continental United States.

To start our assessment of the sensor method we look at basic indicators, such as the entry time, the total number of messages, and the counts of friends and followers for each user. We observe that users with early entry times are characterised by an increased level of activity, as seen in Figure 3A.

Early entrants also have higher network centrality expressed by their in-degree (number of followers) and out-degree (number of friends or followees), shown in Figure 3B. These direct relationships between entry times and other characteristics are especially pronounced in the pre-landing stage of the hurricane, weakening in the post-landing stage.

An example of a distribution of messages over time is shown in Figure 4 for a random control sample and its corresponding sensor group. The inset presents a histogram of tweets, with negligible level of initial activity that builds up and peaks in the landfall day, slowly falling off afterwards. The pattern is largely the same for control and sensor groups, except for the absolute level of activity, with the sensor group being more active. The cumulative distribution function of entry times shows that the sensor group curve is shifted to the left, confirming earlier entry times. The size of the sample in this particular example is 5,000 users, but all sample sizes considered in the study exhibit a similar left shift in the cumulative distribution and an elevated level of activity. The magnitude of the shift and the scale of difference in the activity level both decrease when the size of a sample increases.

Preliminary findings discussed above indicate the link between awareness and centrality, which results in early awareness of sensor groups. It is important to estimate the magnitude of the lead-time and the influence of other factors, in particular the size of the sample and geographical location of users. Lead-time as a function of sample size is presented in Figure 5 (sampling without control over location is given in panel A in a solid black line). In the range of sample sizes considered, the lead-time varies between -11 and -5 hours, with sensor groups consistently showing earlier average entry time. An increase in the size of the sample shortens the lead time and reduces its variance, as previously reported in other studies [32], which is explained by the asymptotic convergence of control and sensor group properties to those of a whole population. Results for key metrics, including tweeting activity levels, lead times and entry times, are summarized in the Supplementary Information, Table S1.

We repeat the analysis with direct control over the location of users. Although it is difficult to accurately identify the area directly affected by the hurricane (because of the multitude of its effects including winds, storm surge, rain, snow, gas and electricity outages), the path and the extent of hurricane force winds provide a good approximation. The track and wind radii data obtained from National Hurricane Center [35] are used to outline the affected area, which is shown in Figure 1. The border of this area is used for selective sampling of individuals directly hit by Hurricane Sandy. Such geographically selective sampling is possible in four combinations: both control and sensor groups are within the affected area; control groups in and sensor groups out; control groups out and sensor groups in; and finally both groups out of the hurricane-affected area.

Key statistics for these combinations are summarised in Figure 5 and Tables S2 – S5 in the Supplementary Information. It can be seen that geography strongly affects awareness. Four combinations of the geographical origin for control and sensor groups all result in the change of the lead-time magnitude compared to the sampling without regard to the location (the solid black line in Figure 5A). The change is moderate if both groups are drawn from the similar geography pool, with affected pairs giving slightly longer and unaffected pairs slightly shorter lead times (see labelled orange and blue trends in Figure 5A). The change is strong for mixed combinations, to the extent that the lead-time reverses its sign (and indicates lagging) when the control group is within and the sensor group is outside of the affected area (green line in Figure 5A). The longest lead times arise for the combination of two factors: geographical relevance and central position in the network topology. This case is illustrated by the purple trend in Figure 5A for control groups formed outside (random position in the topology and low geographical relevance) and sensor groups within the disaster area (high geographical relevance and central position in the network topology). It could be argued that the direct relevance of the event influences one's behaviour in seeking, transmitting and generating information more than one's position as a central node of a social network. A similar explanation is coined to explain other digital traces of Hurricane Sandy, i.e. photographs posted on Flickr [36], where the number of pictures peaks close to the landing time and suggests that observed severity of the disaster may motivate people to document it with higher intensity. Finally, it is noteworthy that the entry times for the sensor groups located inside the affected area are actually negative and correspond to the pre-landing phase of the hurricane, see Table S2 and S4 in the Supplementary Information.

In summary, our experiments show that the sensor method results in the awareness advantage on a scale between 3 and 26 hours, depending on the sample size and geographical origin of the groups.

To evaluate statistical significance of the lead times obtained above, we compare them to the null model where the timestamps of all messages in our database are randomly shuffled. Such a null model preserves the correlation between centrality and normal tweeting frequency, serving as an upper limit on the performance of the sensor method assuming that every user tweets about the disaster shortly after becoming aware about it. Comparison is presented in Figure 5B, with the null model lead times exceeding those in the actual data. This confirms that the spread of the Sandy-related information on Twitter is not purely viral and endogenous, as in that case the actual lead times would outperform the null model [32]. Future development of more complex null models, better suited for exogenous processes, may be required to adequately test experimental results.

## 3.2 Sentiment study

We demonstrated above that the sensor method is generally successful in selecting users with high centrality, activity and awareness. An important question remains if they also differ in their emotional response. To study this, we employ several sentiment analysis techniques. Primarily, we use the sentiment scores generated by a proprietary algorithm from a data provider, analytics company Topsy Labs [37]. During the sentiment analysis of a message each word is matched against a dictionary of keywords and assigned a weight that reflects its emotional impact. Total weights are calculated, normalised by the word count and returned as either a relative sentiment (average of all scores taking into account their sign) or an absolute sentiment (average of absolute values). In our analysis we use the relative sentiment, because it is indicative both of the strength and polarity of sentiment. We also use discrete scores (1, 0 or -1) to distinguish positive, neutral and negative messages and to monitor their fractions in the stream of messages posted.

Since the exact details of the sentiment detection algorithm (i.e. the dictionaries of emotion words and their respective weights) were not published by Topsy and thus cannot be fully reproduced, we verify the analysis with two additional techniques freely available for academic research. The first one is a general-purpose text analysis library Linguistic Inquiry and Word Count [38], which is widely used for detection and classification of emotions in texts. On the most basic level, LIWC provides frequencies of occurrence of positive and negative emotional markers in texts. To combine these two measures into a single metric of sentiment polarity we follow Taboada et.al. [39] and use the difference between the positive score and the scaled (by the factor of 1.5) negative score, a procedure that compensates for a statistical prevalence of positive emotions. The second tool is the SentiStrength by Thelwall et.al. [40], developed specifically for the sentiment classification in short online messages characterised by the frequent use of non-standard spelling, slang, abbreviations and emoticons. Our comparison shows that all three techniques produce highly consistent temporal sentiment trends, shown in Figure 6, that differ only in the scaling factor and in the case of SentiStrength a moderate vertical offset (upward shift of approximately 10% of the peak value is implemented to bring the trend inline with LIWC and Topsy). We conclude that all of these techniques are equally adequate for the study, and the behaviour detected is robust regardless of the specific measurement tool applied.

The temporal evolution of sentiment is tracked as follows: we discretize time into non-overlapping bins of equal duration and take an average of relative sentiment scores for messages posted during each time step. To suppress the noise, we use basic smoothing by a three-point running average, when a value in a time-series is averaged with its nearest neighbours. Typical hourly sentiment trends are shown in Figure 7A-D for the control and sensor groups drawn from various combinations of affected or unaffected areas. The trends are quite noisy, making it necessary to analyse sentiment behaviour over large samples, in this particular instance of 100,000 users. Sentiment features a noticeable diurnal oscillation pattern, previously reported in the analysis of daily and seasonal variations in online activity [41]. Discretising time by days produces smoother curves of a lower temporal resolution, as the ones shown in Figure 7E-H.

Sentiment behaviour exhibits certain general features. First, there is little difference in the dynamics of evolution between the sensor and control groups. Maxima and minima of the control and sensor trends are normally positioned at the same points in time, see for instance Figure 7A and D (similar geography of control and sensor groups). It appears that the awareness advantage of the sensor group does not affect emotional response during the actual disaster itself. While on average sensor users start to monitor the event earlier, sentiment trends suggest that the emotional content of messages is situational, reflects current events, and is universal regardless of the network centrality. Second, being directly affected by the disaster has a detectable effect on the strength of the sentiment. Sample groups formed outside of the disaster zone, regardless of whether they are control or sensor groups, consistently show more positive levels of sentiment illustrated by the vertical shift of corresponding sentiment trends in Figure 7B and C (or Figure 7F and G).

The composition of the message stream evolves with time, as illustrated in Figure 8. The fractions of positive and negative messages are relatively stable, oscillating daily at a certain level. During the disaster phase, the number of negative messages grows at the expense of the positive ones and results in a distinct negative overall sentiment, which lasts approximately from -100 to +100 hours. This increase in the frequency of negative messages at the expense of positive ones potentially gives an opportunity for a simple and universal monitoring technique. Checking the sentiment of a randomly selected sample for the negative overall sentiment, where the share of negatives grows at the expense of positives, may suffice to detect both the occurrence and location of an emergency or disaster. More broadly, the same concept may be applicable to any topic of prominence reflected in the interactions online. For instance, the period of negative sentiment around -300 hours is due to the October 17 presidential debates about the hotly disputed topic of energy policy. The sharp drop in sentiment at +220 hours is due to the weather related tweets discussing the November northeastern storm and the associated snowfall at a time when people still suffered severe consequences from Sandy, including power outages. Notably, this sharp drop is more pronounced in groups drawn from the disaster-affected region.

As an illustration of the sentiment sensing technique, we apply it to the United States in the period between October 21 and November 7. On a regular spatial grid all messages are aggregated hourly and the average sentiment is calculated to obtain a spatio-temporal evolution of density and sentiment of tweets (see Video S1 in the Supplementary Information online). Two snapshots for October 25 and October 29 are presented in Figure 9. The top panel shows a distribution of messages on October 25, when activity is low and the sentiment is mostly neutral or positive, with the exception of the Miami area. At this stage, Sandy has just passed Jamaica and Cuba, and Florida is directly under threat, which contributes to the "negativity" cluster in Miami. Closer to the landing time (bottom panel) activity increases significantly and the sentiment in the affected area is overwhelmingly negative. Interestingly, regions unaffected by the hurricane still demonstrate rather neutral reaction, except for the major urban centres.

## 4 Discussion

In this empirical study we found that the method based on the "friendship paradox" is generally successful in forming sensor groups with an awareness advantage over the randomly selected control groups. The magnitude of the lead-time varies with the size of a sample, showing an advantage of up to 11 hours in small samples of 500 users. This advantage shortens to 5 hours when the size increases to 100,000 users. Lead-time can change significantly when geographical restrictions are imposed on the formation of control and sensor groups, especially if one of them is from the disaster-affected area and the other is not. Maximum advantage detected in our study was about 26 hours and resulted from the combination of high network centrality and geographical relevance.

Additional study of sentiment revealed that the emotional response was universal and followed a similar temporal evolution pattern in both control and sensor groups. The stream of messages changed its composition during the active phase of the disaster, and the increased fraction of negative messages pushed the average sentiment into negativity. Similar behaviour was observed on the shorter scale during the observation period and was linked to other prominent events (the presidential debates and the northeastern snowstorm). Features demonstrated by the sentiment are promising in terms of developing a universal sensing technique that does not require any preconditioning in the form of specific keywords to monitor.

Our study presents a first empirical investigation of the sensor method in a network where information propagates in a mixed mode, both endogenously and exogenously, and factors like a relatively short time scale and strong geographical nature of a disaster affect performance of the method. The lead times we obtained may be sufficient for individuals to improve their own preparedness to a threat like a hurricane (warn others, stockpile water, food, medicines, fuel, batteries, protect properties, etc.), but unlikely to give authorities enough time to adjust their global large-scale response. Nonetheless, the importance of the efficient pathway for the propagation of emergency information, provided by sensor groups, should not be underestimated. Early exposure to witness reports is a factor in compliance with authorities, because behaviour in disasters is often collective [42] and evacuation decisions depend heavily on the perception of risk and peer influence [43]. There is also an inherent resistance towards recognising potential dangers [44] and online interactions may either facilitate better response to threats or create undesired consequences, like panic [45,46]. Management of the information flow must therefore be included into communications part of emergency planning guidelines [47].

It should be understood that the specific nature of a threat would be reflected in the performance of the sensor method. Since hurricanes, compared to other disasters, evolve slowly and are characterised by exceptional predictability using modern atmospheric simulation techniques, the value of the advantage on the scale of hours may be questionable as an early warning. But this might not be the case in other scenarios without extended warning time and predictability: for instance earthquakes [17-21], terrorist attacks, technological catastrophes, forest fires and flash floods. In events like these, a social network may effectively serve as a primary source of information from eyewitnesses [48], as well as a medium for its distribution.

The sentiment sensing technique proposed here is an attractive alternative to the existing methods of disaster detection on Twitter [20,21], which are based on the monitoring of message frequency and text mining for event-specific keywords. Our approach is indifferent to the nature of a disaster and does not require any filtering of a message stream to extract event-related tweets. However, additional study is needed to evaluate how well such a technique performs in an unfiltered stream of information.

One important limitation to the reliability of quantitative estimates of the lead-time via digital traces on social media lies in the assumption of a direct correlation between awareness and online activity on the topic. Such correlation needs additional experimental validation by traditional sociological methods. Regional and demographic differences are likely to exist in online behaviour, based on the communication platform adoption rate and patterns of use across groups from different regions, age, socio-economic status or cultural heritage. However, rigorous validation of these assumptions is impossible on the basis of the data that we use and is therefore beyond the scope of our research.

Overall, our findings confirm the potential of sensor method for efficient early detection of emergency information and offer new sentiment sensing technique for detection and localisation of disasters.

# 5  Acknowledgements

Yury Kryvasheyeu, Haohui Chen, Pascal Van Hentenryck and Manuel Cebrian acknowledge the support of Australian Government represented by the Department of Broadband, Communications and Digital Economy and Australian Research Council through the ICT Centre of Excellence program.

# 6  Author contribution statement

The authors contributed equally to the analysis performed, interpretation of results and preparation of the manuscript.

# 7  Additional information

## 7.1  Competing financial interests

The authors declare no competing financial interests.

## 7.2  Data accessibility

The data was obtained through the analytics company Topsy Labs, and due to Twitter's policy is not available for re-distribution. We verified the results of the proprietary sentiment classification algorithm by Topsy with other openly available tools (LIWC, SentiStrength) to guarantee reproducibility of our findings. All control and sensor samples, which contain a limited subset of information required for aggregated analysis without user-identifying details, may be provided upon request.

# 9 Figures captions and legends

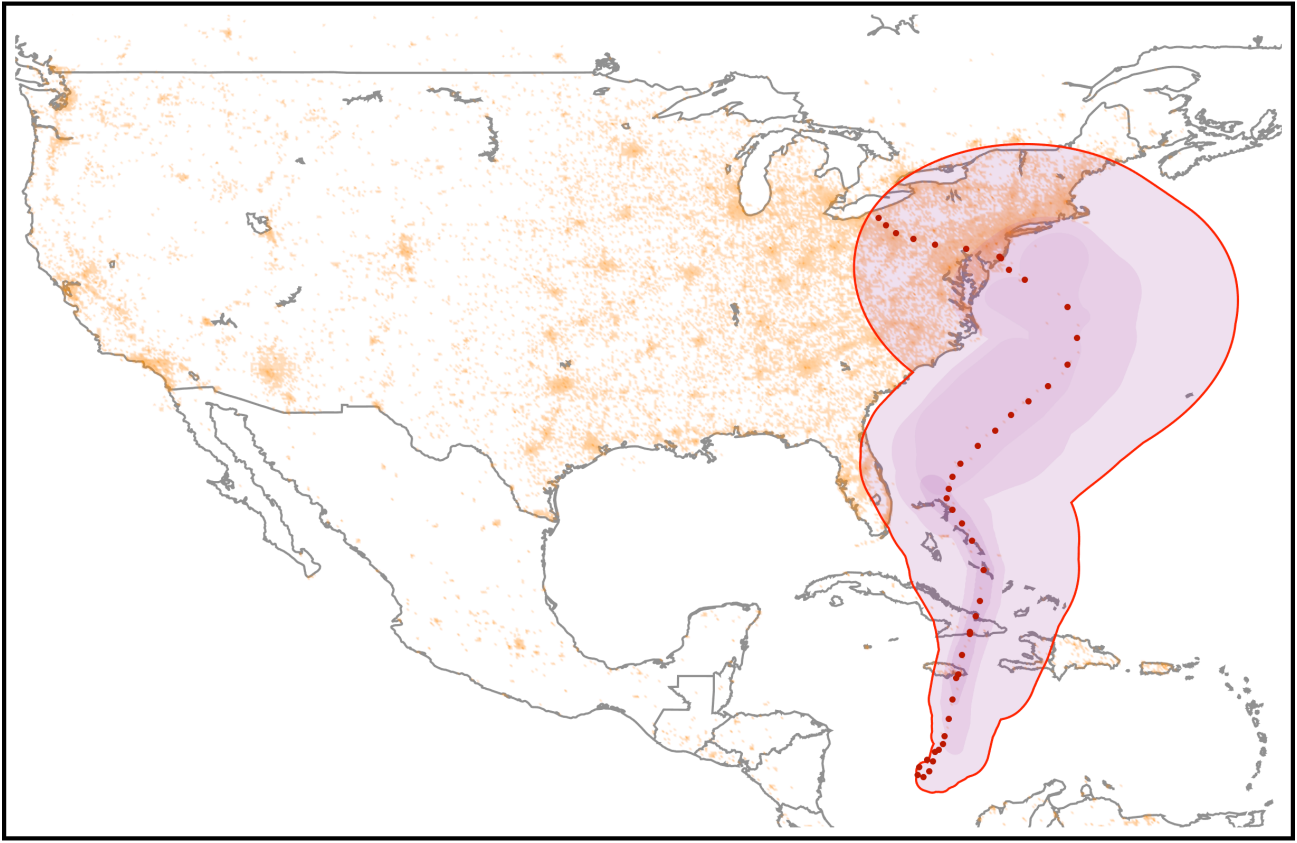

**Figure 1 The track and approximate extent of Hurricane Sandy, combined with the heatmap of geolocated tweets density.**

The path of the hurricane from the moment of its formation until dissipation is accompanied by the approximate extent of the hurricane force winds. Three threshold levels distinguished by shading correspond to the hurricane forces between Categories 1 and 3 (34, 50 and 64 knots respectively). An outer extent of the Category 1 winds is outlined in red and serves as a border of the area directly affected by the hurricane.

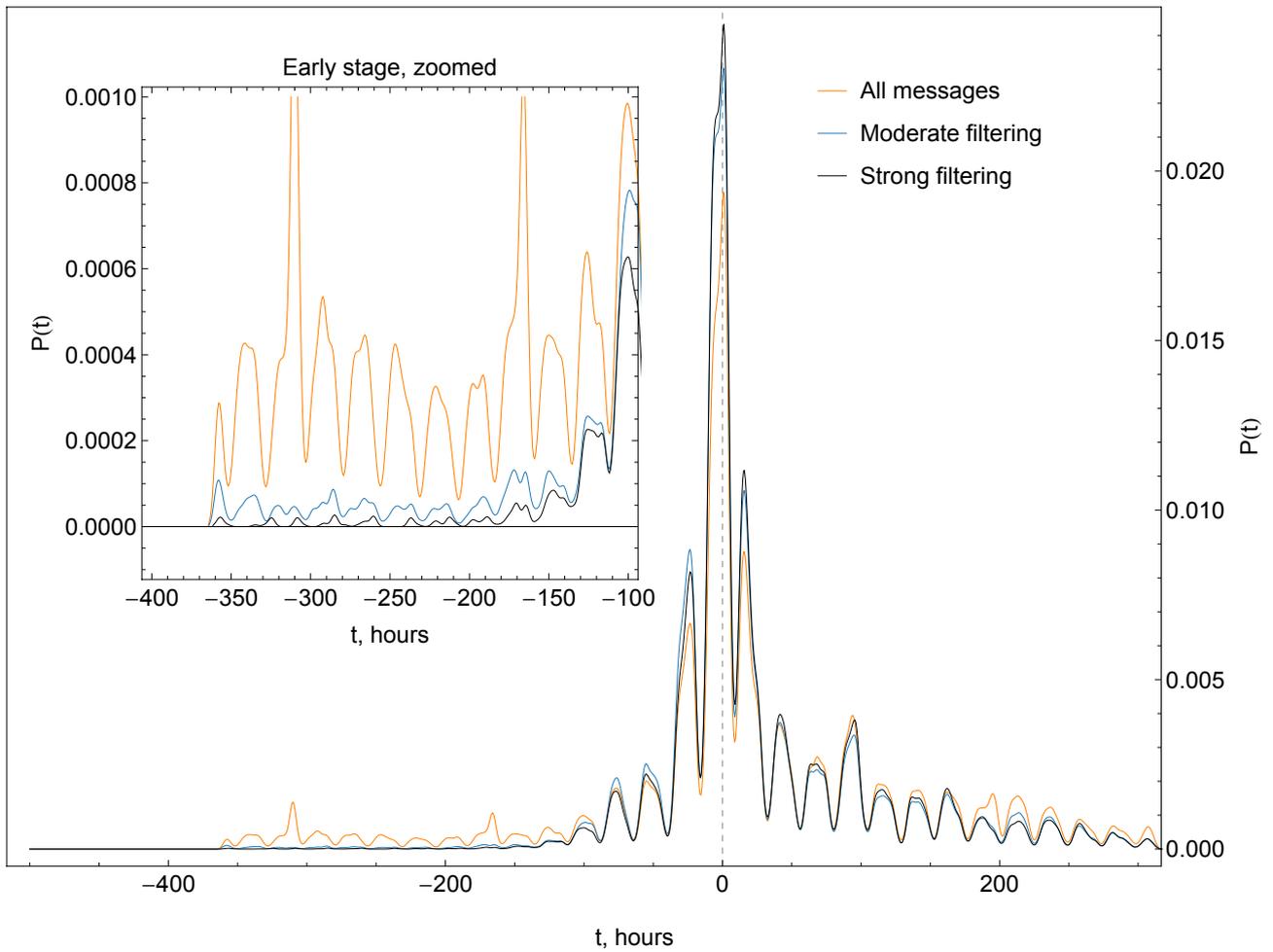

**Figure 2 Effect of different levels of content filtering.**

Strong filtering avoids early "noise" of irrelevant messages that may skew the estimate of an entry time. The most reliable form of such filtering is achieved only by including those messages that have "sandy" as a part of a text or as a hashtag.

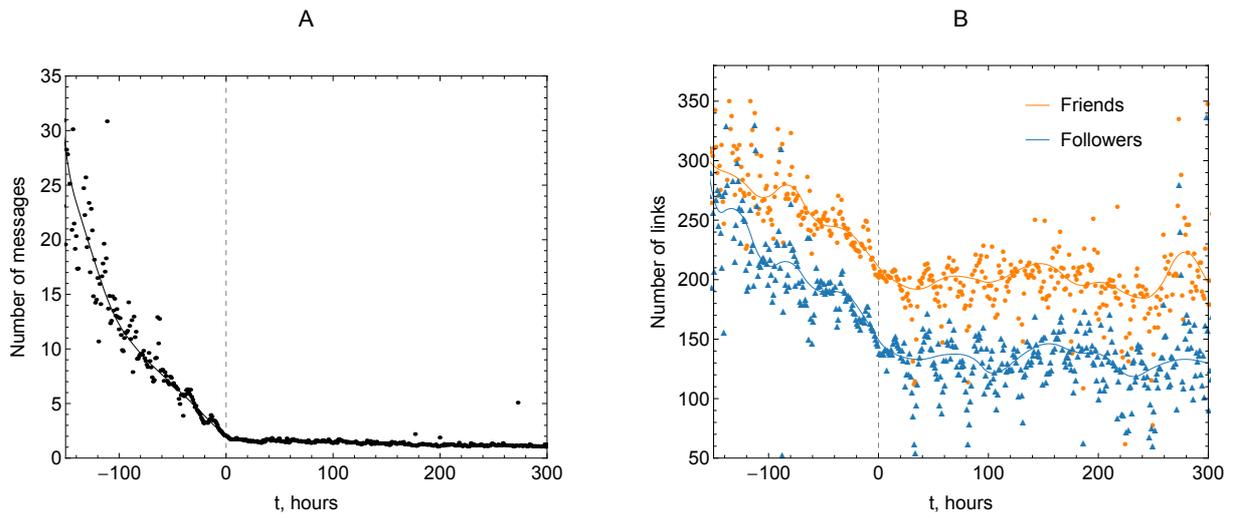

**Figure 3 Average activity (A) and mean number of friends and followers (B) for users as a function of their entry time.**

Analysis shows that users who appear in the dataset early demonstrate higher level of activity and are characterized by the higher counts of friends and followers (occupy a more central network position). These features are especially pronounced in the pre-landing stage of the hurricane history (landing time is taken as a reference zero point). The fact that both activity and centrality correlate with entry time (awareness) suggests that the "friendship paradox" holds and sensor groups have an advantage of awareness lead-time, the magnitude of which is to be established.

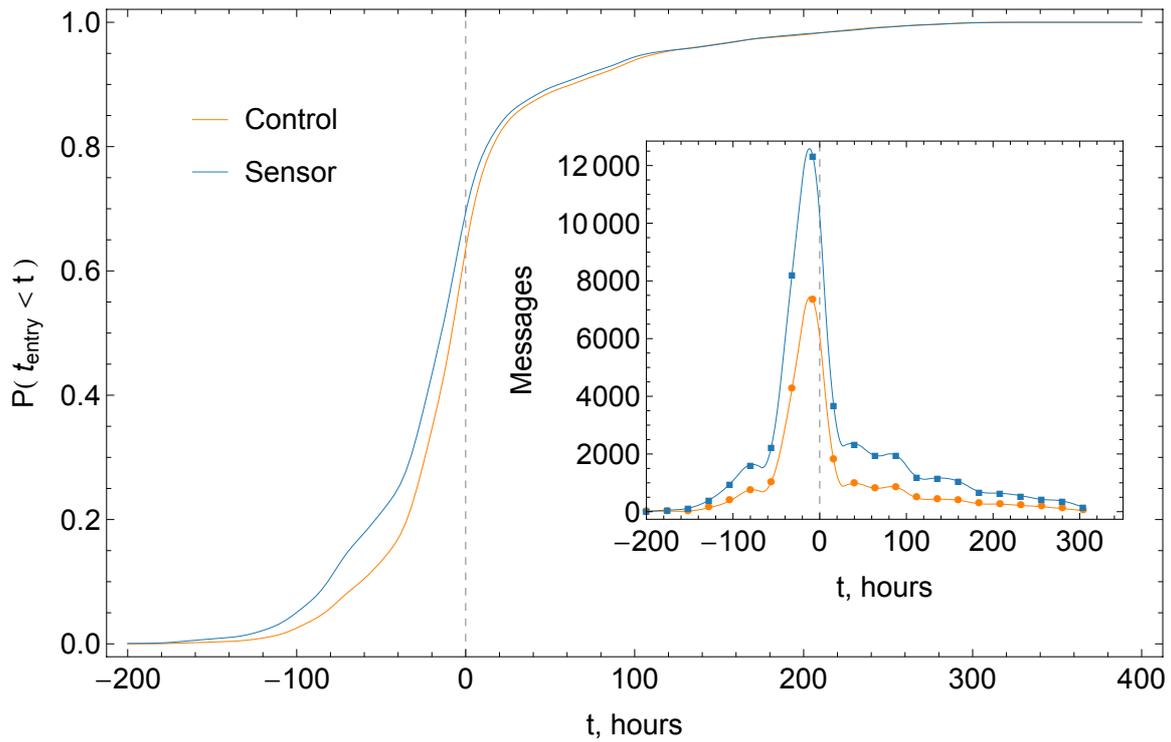

**Figure 4 Typical cumulative distribution functions of messages posted by a random control group and its corresponding sensor group.**

Figure shows a distribution of messages over time, discretised on the daily basis with diurnal oscillations (and discrete steps in cumulative representation) smoothed out by the Gaussian kernel density estimation with 8-hour bandwidth. The inset shows simple daily count histograms, which peak in the landing day, with the sensor group activity at significantly higher level. Left shift of the sensor group's cumulative distribution confirms that it consistently leads in terms of the entry time (awareness).

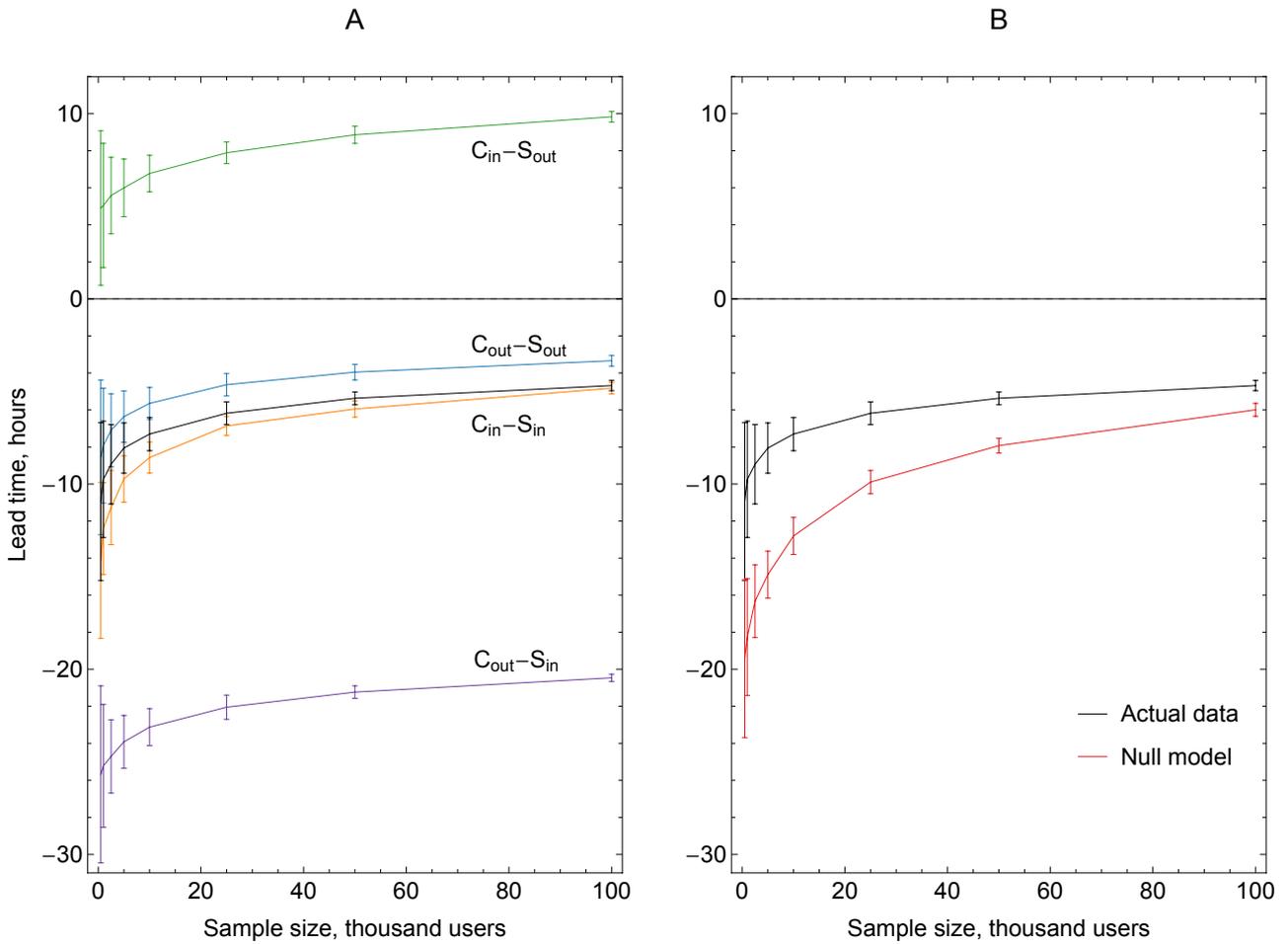

**Figure 5 Lead-time magnitude for control and sensor groups of various combinations of geographical origin (A) and comparison against a Null model with shuffled timestamps (B).**

The lead-time magnitude and its variance (A) both decrease with an increasing sample size. Geography plays an important role in determining the scale of the awareness advantage. Longest lead-time is achieved through the combination of network centrality and geographical relevance in the case of "control out - sensor in" combination. Geographical factor outweighs the network effect, as illustrated by the positive lead times (or rather lag times in this case) for the "control in - sensor out" combination. Relative underperformance of the actual data against a randomised null model (B) may be caused by the exogenous, rather than endogenous, mode of information spread and the correlation between centrality and tweeting frequency.

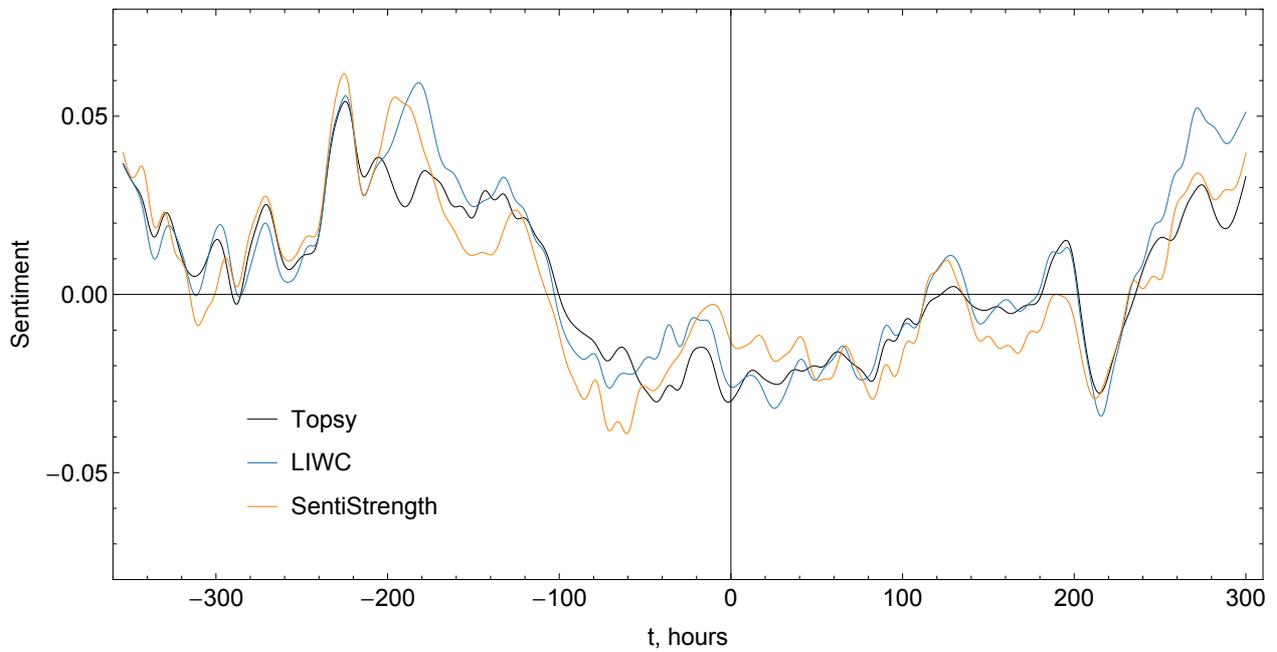

**Figure 6 Temporal evolution of sentiment measured by three different techniques (Topsy, LIWC and SentiStrength).**

Comparison of the proprietary Topsy algorithm against freely available alternatives shows that sentiment trends are consistent between different tools. Positive (*posemo*) and negative (*negemo*) emotion scores provided by Linguistic Inquiry and Word Count library were combined into a single polarity measure as $score = posemo - 1.5 \cdot negemo$, and the result of SentiStrength was shifted upwards by approximately 10% of its peak value.

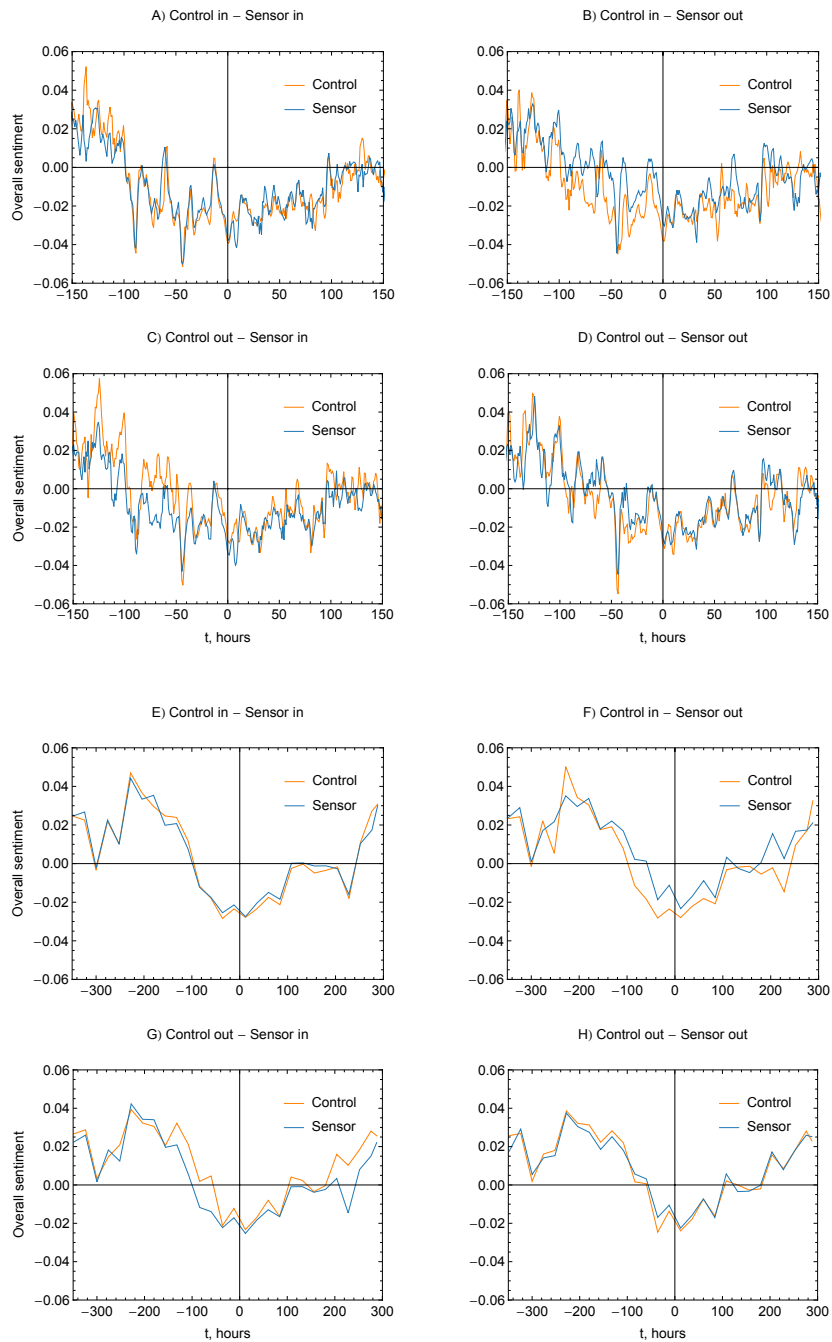

**Figure 7 Hourly (A-D) and daily (E-F) sentiment trends for control and sensor groups of different geography, as identified in the panel titles.**

Sentiment trends exhibit high level of random noise when aggregated on the basis of short time steps (A-D). If averaged over a longer period (E-F), an overall positive baseline level appears, which temporarily drops into negativity in the lead up to and the aftermath of the Hurricane Sandy landing (approximately -100 to +100 hours). There is no discernible difference, i.e. detectable horizontal shift, in the temporal evolution of sentiment between control and sensor groups, suggesting that the emotional response is universal.

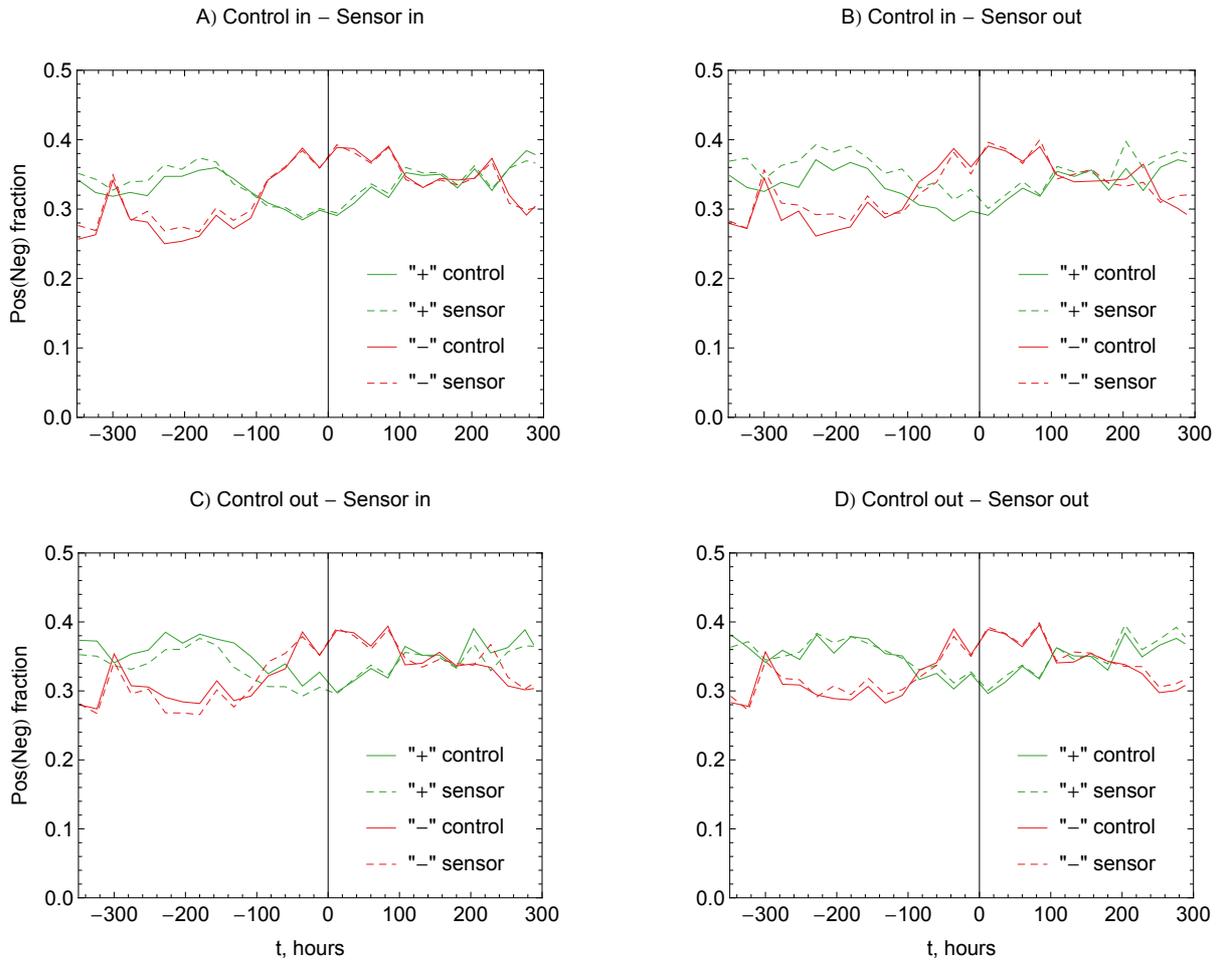

**Figure 8 Daily trends in the composition of the message stream.**

We monitor the fraction of positive (solid green line for control and dashed green for sensor groups) and negative messages (solid red for control and dashed red for sensor groups) in the total volume of all tweets. During the most severe stage of the hurricane and after its landing the composition undergoes transition from predominantly positive to predominantly negative.

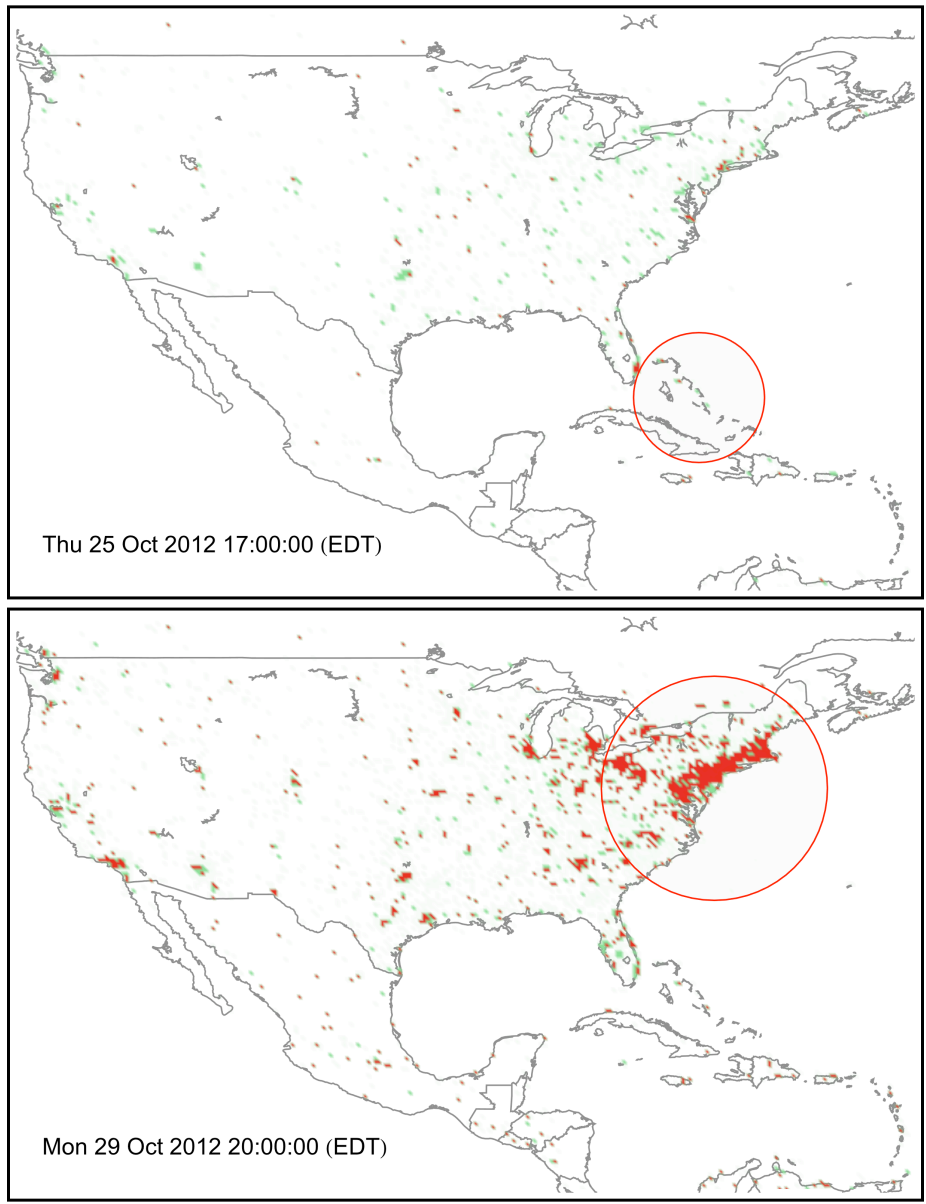

**Figure 9 Comparison of density and sentiment for messages posted during an hour-long period at 17:00 EDT October 25 (top) and 20:00 EDT October 29 (bottom).**

The density and polarity of sentiment is highlighted by green (positive) or red (negative). At the early stage, prevalent sentiment is either neutral or positive, and the interest in the hurricane is comparatively low, except for the Miami area. Close to the landing time (bottom) the sentiment in the area affected by the hurricane is overwhelmingly negative.

# Performance of Social Network Sensors During Hurricane Sandy

Yury Kryvasheyeu[1], Haohui Chen[1], Esteban Moro[2,3], Pascal Van Hentenryck[1,4], and Manuel Cebrian[1,*]

[1]*National Information and Communications Technology Australia, University of Melbourne, Victoria 3010, Australia*

[2]*Department of Mathematics & GISC, Universidad Carlos III de Madrid, Madrid, Leganés 28911, Spain*

[3]*Instituto de Ingeniería del Conocimiento, Universidad Autónoma de Madrid, Madrid 28049, Spain*

[4]*Research School of Computer Science, Australian National University, ACT 0200, Australia*

*\* Corresponding Author: MC, manuel.cebrian@nicta.com.au*

## Supplementary Information

Contains five tables and two figures referenced from the main text:

1. Tables S1 – S5, with summary of lead times and other metrics for all samples of various sizes and different combinations of geographical origin for control and sensor groups
2. Table S6, with a list of keywords in messages included into extended dataset
3. Figures S1 and S2, with histograms of use for specific keywords to illustrate relevance filtering

Also, the Video S1 illustrating the sentiment sensing would be uploaded and available online.

**Table S1** Average activities $N_C$ and $N_S$ (messages per user), entry times $\langle t_C \rangle$ and $\langle t_S \rangle$, and lead times $\Delta t$ (in hours) for control and sensor groups formed without location restrictions.

| Sample size | $N_C$ | $N_S$ | $\Delta t \pm \sigma(\Delta t)$, h | $\langle t_C \rangle$, h | $\langle t_S \rangle$, h |
|---|---|---|---|---|---|
| 500 | 3.23 | 8.65 | -10.9 ± 4.26 | 11.2 | 0.25 |
| 1000 | 3.17 | 8.07 | -9.74 ± 3.14 | 10.7 | 0.99 |
| 2500 | 3.24 | 7.34 | -8.93 ± 2.14 | 10.7 | 1.81 |
| 5000 | 3.20 | 6.63 | -8.06 ± 1.36 | 10.7 | 2.61 |
| 10000 | 3.20 | 5.97 | -7.30 ± 0.90 | 10.8 | 3.51 |
| 25000 | 3.22 | 5.13 | -6.18 ± 0.61 | 10.8 | 4.63 |
| 50000 | 3.20 | 4.61 | -5.37 ± 0.34 | 10.8 | 5.42 |
| 100000 | 3.21 | 4.16 | -4.68 ± 0.27 | 10.9 | 6.21 |

**Table S2** Average activities $N_C$ and $N_S$ (messages per user), entry times $\langle t_C \rangle$ and $\langle t_S \rangle$, and lead times $\Delta t$ (in hours), with both control and sensor groups formed from users affected by the hurricane: "Control In – Sensor In" sampling.

| Sample size | $N_C$ | $N_S$ | $\Delta t \pm \sigma(\Delta t)$, h | $\langle t_C \rangle$, h | $\langle t_S \rangle$, h |
|---|---|---|---|---|---|
| 500 | 4.10 | 11.4 | -13.8 ± 4.29 | -0.07 | -13.9 |
| 1000 | 4.13 | 10.3 | -12.8 ± 2.64 | -0.18 | -12.9 |
| 2500 | 4.10 | 8.87 | -11.3 ± 1.86 | 0.08 | -11.2 |
| 5000 | 4.13 | 8.01 | -9.80 ± 1.31 | -0.05 | -9.85 |
| 10000 | 4.12 | 7.03 | -8.53 ± 0.92 | -0.05 | -8.59 |
| 25000 | 4.12 | 6.05 | -7.08 ± 0.60 | 0.02 | -7.06 |
| 50000 | 4.11 | 5.45 | -5.90 ± 0.36 | -0.03 | -5.92 |
| 100000 | 4.10 | 5.00 | -4.85 ± 0.25 | 0.03 | -4.83 |

**Table S3** Average activities $N_C$ and $N_S$ (messages per user), entry times $\langle t_C \rangle$ and $\langle t_S \rangle$, and lead times $\Delta t$ (in hours) for control groups affected and sensors unaffected by the hurricane: "Control In – Sensor Out" sampling.

| Sample size | $N_C$ | $N_S$ | $\Delta t \pm \sigma(\Delta t)$, h | $\langle t_C \rangle$, h | $\langle t_S \rangle$, h |
|---|---|---|---|---|---|
| 500 | 4.07 | 7.73 | 5.14 ± 4.29 | 0.08 | 5.21 |
| 1000 | 4.15 | 7.43 | 5.15 ± 3.23 | 0.12 | 5.27 |
| 2500 | 4.10 | 6.78 | 6.07 ± 1.88 | 0.05 | 6.12 |
| 5000 | 4.10 | 6.28 | 6.37 ± 1.38 | -0.03 | 6.33 |
| 10000 | 4.11 | 5.74 | 6.88 ± 1.06 | -0.08 | 6.79 |
| 25000 | 4.11 | 4.98 | 8.03 ± 0.65 | -0.00 | 8.03 |
| 50000 | 4.11 | 4.48 | 8.98 ± 0.50 | -0.01 | 8.97 |
| 100000 | 4.11 | 4.03 | 10.1 ± 0.31 | -0.04 | 10.0 |

**Table S4** Average activities $N_C$ and $N_S$ (messages per user), entry times $\langle t_C \rangle$ and $\langle t_S \rangle$, and lead times $\Delta t$ (in hours) for control groups unaffected and sensors affected by the hurricane: "Control Out – Sensor In" sampling.

| Sample size | $N_C$ | $N_S$ | $\Delta t \pm \sigma(\Delta t)$, h | $\langle t_C \rangle$, h | $\langle t_S \rangle$, h |
|---|---|---|---|---|---|
| 500 | 2.86 | 11.9 | -26.2 ± 4.08 | 15.8 | -10.4 |
| 1000 | 2.87 | 10.9 | -25.7 ± 3.16 | 16.0 | -9.69 |
| 2500 | 2.91 | 9.88 | -24.9 ± 2.06 | 15.8 | -9.06 |
| 5000 | 2.85 | 9.03 | -24.2 ± 1.34 | 16.0 | -8.22 |
| 10000 | 2.87 | 8.19 | -23.2 ± 1.01 | 15.8 | -7.44 |
| 25000 | 2.86 | 7.05 | -22.1 ± 0.53 | 15.8 | -6.31 |
| 50000 | 2.87 | 6.34 | -21.3 ± 0.36 | 15.8 | -5.52 |
| 100000 | 2.87 | 5.66 | -20.4 ± 0.28 | 15.8 | -4.67 |

**Table S5** Average activities $N_C$ and $N_S$ (messages per user), entry times $\langle t_C \rangle$ and $\langle t_S \rangle$, and lead times $\Delta t$ (in hours) for both groups unaffected: "Control Out – Sensor Out" sampling.

| Sample size | $N_C$ | $N_S$ | $\Delta t \pm \sigma(\Delta t)$, h | $\langle t_C \rangle$, h | $\langle t_S \rangle$, h |
|---|---|---|---|---|---|
| 500 | 2.88 | 6.88 | -8.14 ± 4.36 | 15.7 | 7.52 |
| 1000 | 2.86 | 6.64 | -7.95 ± 3.05 | 15.9 | 7.91 |
| 2500 | 2.87 | 6.09 | -7.23 ± 2.05 | 15.8 | 8.59 |
| 5000 | 2.87 | 5.59 | -6.23 ± 1.41 | 15.6 | 9.40 |
| 10000 | 2.89 | 5.03 | -5.46 ± 0.98 | 15.7 | 10.3 |
| 25000 | 2.86 | 4.38 | -4.57 ± 0.61 | 15.8 | 11.3 |
| 50000 | 2.87 | 3.97 | -3.97 ± 0.38 | 15.8 | 11.8 |
| 100000 | 2.86 | 3.61 | -3.36 ± 0.25 | 15.8 | 12.5 |

**Table S6** Keywords used to form the extended dataset with their corresponding message count.

| Keyword | Count | Keyword | Count | Keyword | Count |
|---|---|---|---|---|---|
| sand | 4846422 | blackout | 213520 | gás | 18818 |
| power | 4825717 | franken | 210277 | wallst | 18423 |
| sandy | 4745099 | mta | 206504 | nopower | 13660 |
| hurricane | 4680290 | frankenstorm | 205467 | stock exchange | 11840 |
| weather | 3333025 | NewYork | 195078 | Con Edison | 11321 |
| storm | 2555196 | nyc marathon | 102838 | comfortablysmug | 9963 |
| New York | 2348535 | Cuomo | 92014 | 911buff | 9521 |
| gas | 1991524 | prayforusa | 91293 | wallstreet | 8855 |
| hurricane | 1906749 | superstorm | 70274 | new-york | 8691 |
| hurricanesandy | 518492 | nyse | 69213 | opsafe | 7714 |
| Governor | 498135 | ConEd | 47749 | Governor Cuomo | 6786 |
| stay safe | 484732 | huracan | 40357 | sandy aid | 3048 |
| recovery | 431591 | wtc | 38567 | pray for usa | 1891 |
| climate | 420217 | climatechange | 38365 | operation safe | 1482 |
| huracán | 371157 | staysafe | 38094 | trading floor | 1130 |
| FEMA | 329789 | ConEdison | 30080 | sandy pets | 929 |
| flooding | 264132 | conedison | 30080 | StockExchange | 181 |
| no power | 261998 | sandypets | 28300 | op safe | 61 |
| climate change | 236009 | nycmarathon | 26179 | GovernorCuomo | 59 |
| wall st | 233411 | sandyaid | 19897 | TradingFloor | 21 |

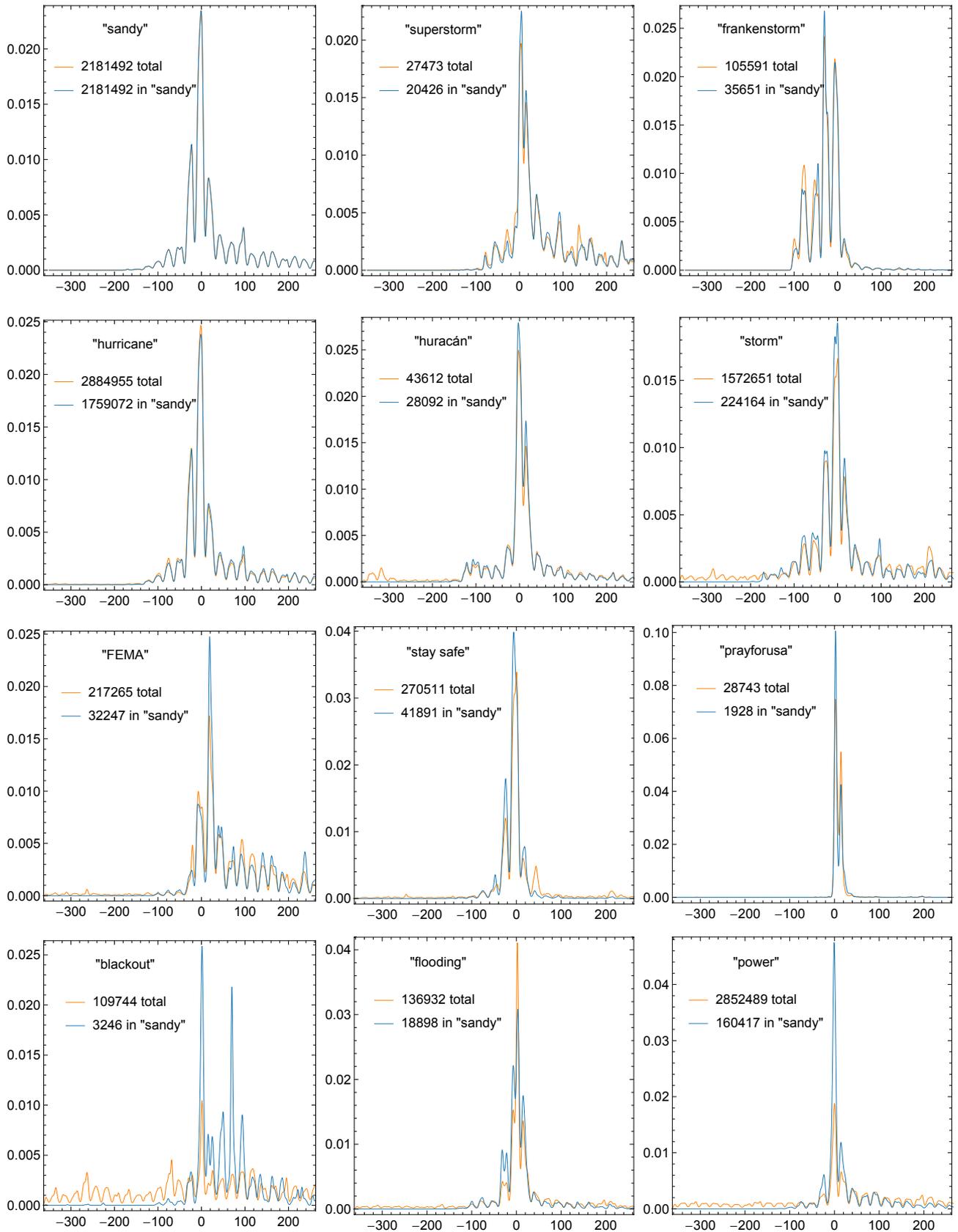

**Figure S1 Histograms of occurrence for messages with specific keywords.** The figure shows probability density functions for the occurrence of a keyword on its own in orange, and in combination with "sandy" in blue. The messages that occur before 22$^{nd}$ October 2012 (approximately -200 hours on horizontal axis) are likely to be irrelevant to Hurricane Sandy and should be filtered out. Histograms are arranged in the approximate order of relevance.

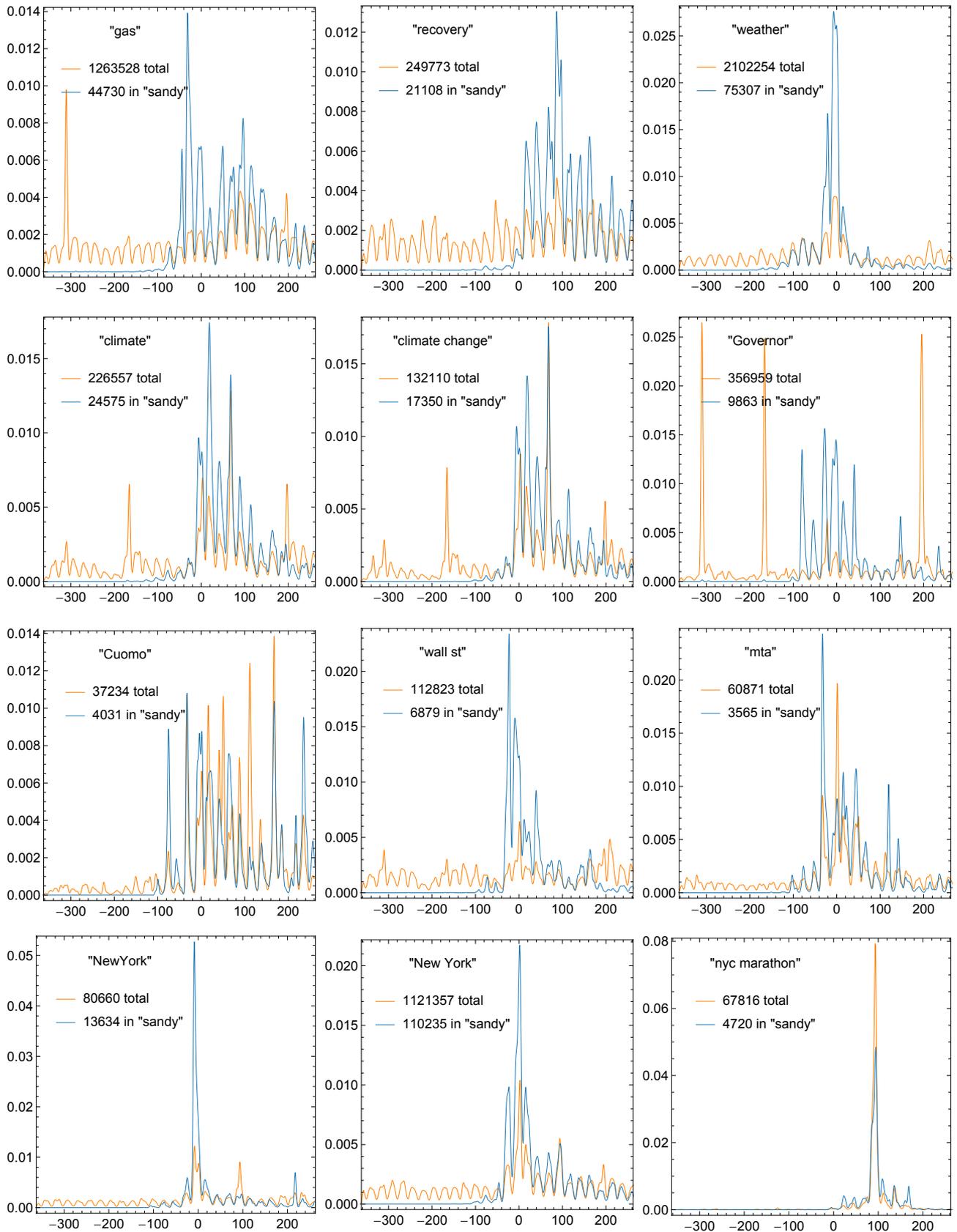

**Figure S2 More keyword occurrence histograms, continued from Figure S1.** This figure continues the sequence from Figure S1, with decreasing level of keyword relevance. Note the frequent incidence of use before 22$^{nd}$ October 2012.